# Prevalence of oxygen defects in an in-plane anisotropic transition metal dichalcogenide

Ryan Plumadore[1], Mehmet Baskurt[5], Justin Boddison-Chouinard[1], Gregory Lopinski[2], Mohsen Modaresi[3], Pawel Potasz[4], Pawel Hawrylak[1], Hasan Sahin[5], Francois M. Peeters[6], Adina Luican-Mayer[1*]

[1]*Department of Physics, University of Ottawa, Ottawa, Canada*
[2]*National Research Council, Ottawa, Canada*
[3]*Department of Physics, Ferdowsi University of Mashhad, Mashhad, Iran*
[4]*Departmet of Theoretical Physics, Wroclaw University of Science and Technology, Wroclaw, Poland*
[5]*Deperatment of Photonics, Izmir Institute of Technology, 35430, Urla, Izmir, Turkey*
[6]*Department of Physics, University of Antwerp, Groenenborgerlaan 171
B-2020 Antwerpen, Belgium*

luican-mayer@uottawa.ca

**Atomic scale defects in semiconductors enable their technological applications and realization of novel quantum states. Using scanning tunneling microscopy and spectroscopy complemented by ab-initio calculations we determine the nature of defects in the anisotropic van der Waals layered semiconductor $ReS_2$. We demonstrate the in-plane anisotropy of the lattice by directly visualizing chains of rhenium atoms forming diamond-shaped clusters. Using scanning tunneling spectroscopy we measure the semiconducting gap in the density of states. We reveal the presence of lattice defects and by comparison of their topographic and spectroscopic signatures with ab initio calculations we determine their origin as oxygen atoms absorbed at lattice point defect sites. These results provide an atomic-scale view into the semiconducting transition metal dichalcogenides, paving the way toward understanding and engineering their properties.**

Intrinsic defects, such as grain boundaries, vacancies, adatoms, and substitutional impurities are ubiquitously present in graphene and transition metal dichalcogenides (TMDs)[1-4]. Harnessing the properties of TMDs for atomic scale opto-electronic devices requires detailed understanding of the nature and effects of structural lattice defects on the material properties. Defects can influence the performance of devices through the formation of electronic states within the semiconducting gap [5,6], can modify the excitonic transitions and photoluminescence response[7], can behave as atom-like quantum emitters[8,9] and can affect the catalytic properties[10]. Although imaging techniques visualized surface defects of bulk TMDs[11-13], identifying their nature and correlating their presence with properties of the material remains challenging. For example, one of the most common imaging techniques, electron microscopy, can result in knock-on sputtering of chalcogen atoms[13], making it impossible to distinguish the intrinsic defects. Moreover, in the case of some TMDs like ReS$_2$, the larger atomic number of the metal can prevent identifying the nature of defects. Therefore, non-invasive techniques like scanning tunneling microscopy and spectroscopy (STM/STS) can be invaluable in imaging and determining the properties of lattice defects.

Here, we use a combination of STM/STS (Figure 1(a)) in tandem with ab initio calculations to determine the structural and electronic properties of ReS$_2$ and its defects. These experimental and theoretical results indicate that a common defect corresponds to oxygen atoms adsorbed at lattice point defect sites.

While most van der Waals (vdW) layered materials are isotropic within the plane, a number of them have lower in-plane crystal symmetry[14]. In-plane anisotropy in layered vdW materials gives additional functionality that has only just began to be explored in black phosphorus (BP)[15], transition metal trichalcogenides (TiS$_3$)[16], rhenium disulfide (ReS$_2$) and rhenium diselenide (ReSe$_2$)[17]. Among them, ReS$_2$ attracted attention as a material that is stable under ambient conditions, with structural anisotropy that translates into further tunability in its electronic [18,19], optical [20-23] and mechanical[24] properties. Structurally, ReS$_2$ is a distorted 1T structure which belongs to the $P\bar{1}$ space group [25,26]. The additional electron in the d orbital of the rhenium atoms favors the existence of metallic Re—Re bonds, responsible for the formation of a superlattice of chains with diamond clusters of 4 Re atoms, as shown schematically in Figure 1(b). This structure was previously visualized by X-ray spectroscopy[25,27] and electron microscopy[13]. Early scanning probe microscopies provided also indications of a Re chain structure [28-30]. When cleaved, the crystal orientation of the ReS$_2$ flakes is apparent, as they maintain the edges parallel to the rhenium chains. This feature is useful for identifying the orientation of a crystal for example in device fabrication [18,31]. Indeed, in the large area STM topographic images, as presented in Figure 1(c), we find that terraces on the surface of ReS$_2$ follow the crystallographic directions with an angle $\gamma \approx$ 120° between the $a$ and $b$ directions. The terraces and steps in Figure 1(c) correspond to single or double atomic layers, with their height measured as shown in the inset.

In higher resolution images, presented in Figure 1(d), we observe the characteristic anisotropic lattice structure with elongated chains. One might be tempted to attribute the measured features to the topmost layer of sulfur (S) atoms[30] which is closest to the tip. However, a more accurate interpretation reveals that the contrast in the atomically resolved STM topographic images originates from local density of states (LDOS). Therefore, the complex dependence of the partial density of states (PDOS) at the surface layer on the energy at which the tunneling process occurs must be considered. The calculated PDOS using ab initio theory is shown in Figure 1(e). The density of states is dominated by the Rhenium-d orbitals, which is related to the flat band dispersion around the top of the valence band (VB) at the Γ point. At higher energies, band crossings lead to a substructure in the density of states. We find that within the energy range of our experiment the Rhenium d-orbitals (in red in Figure 1(e)) are responsible for the largest contribution to the density of states. This implies that the metal atoms are predominantly involved in the tunneling process and therefore provide the contrast in the atomically resolved topographic images. This is further confirmed by simulated images of the STM measurement in Figure 1(f).

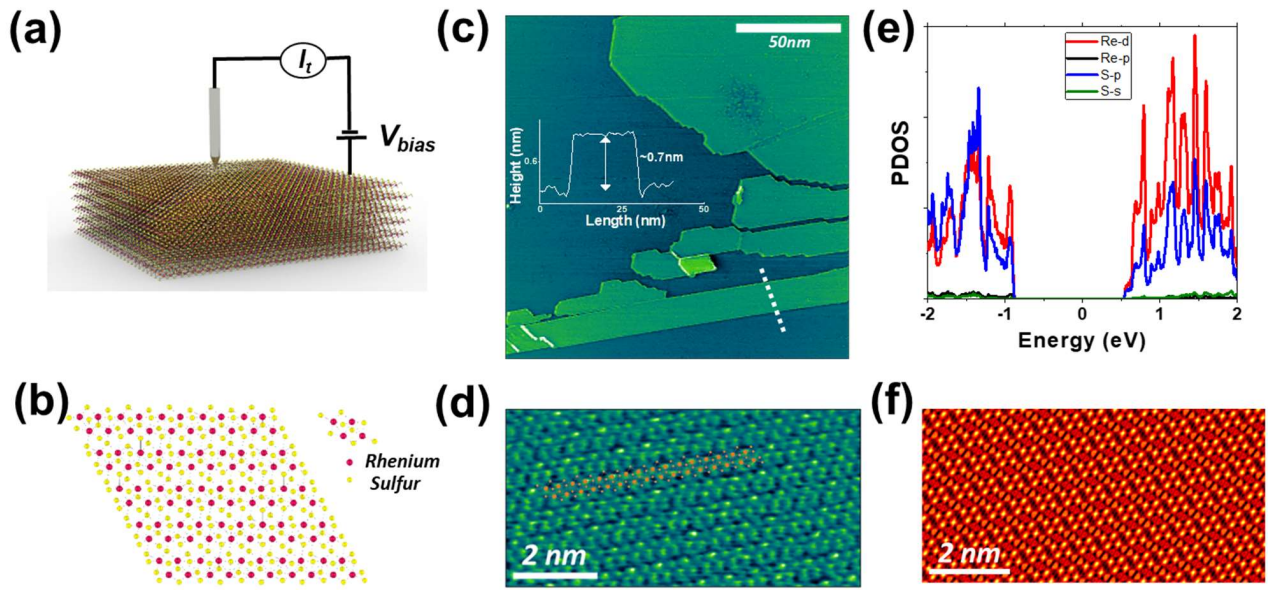

**Figure 1: (a) Schematic of the STM experiment (b) Top and side view of the ReS$_2$ structure. (c) STM topographic map of step edges on the surface of a ReS$_2$ crystal ($V_b$= -1.60V; $I_T$ = 450pA). Inset: Height profile along the white dotted line. (d) STM topographic image highlighting the Re chains with a spheres model as a guide to the eye. (e) Calculated PDOS of monolayer ReS$_2$. (f) Simulated STM image of the ReS$_2$ lattice.**

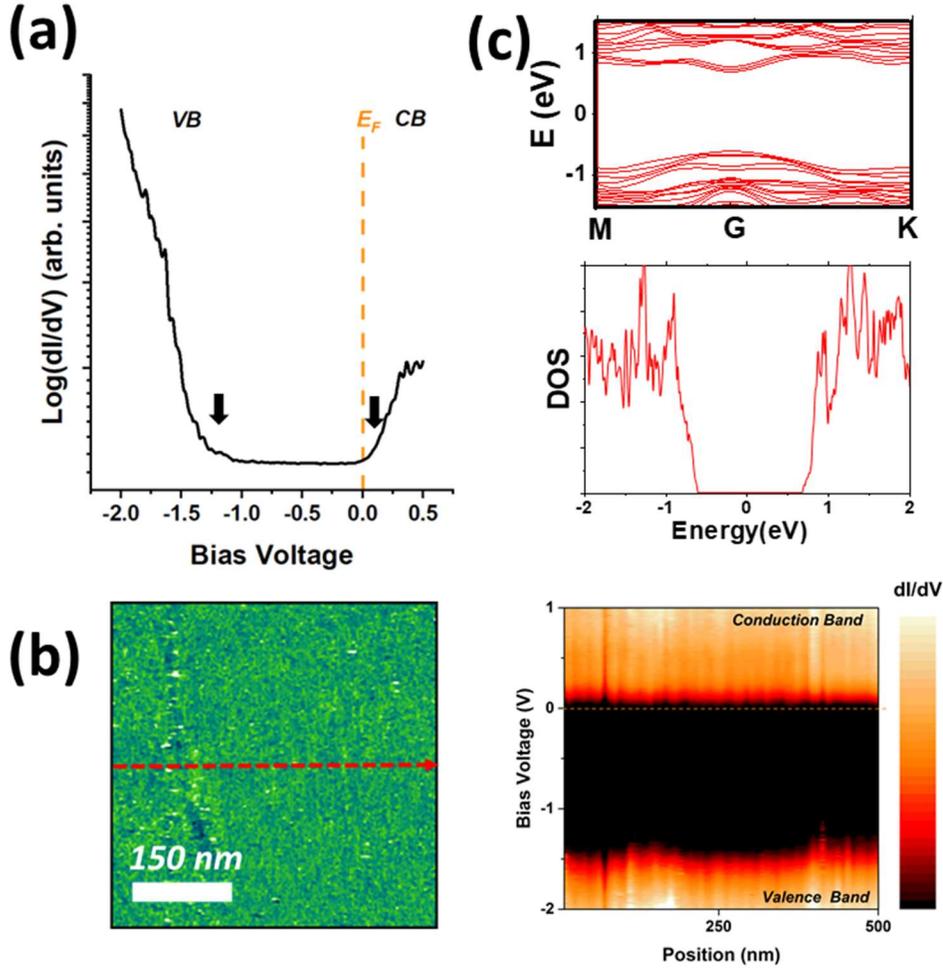

**Figure 2: (a) Scanning Tunneling Spectroscopy averaged on the surface of ReS$_2$. The logarithmic scale was used to highlight valence band (VB) and, conduction band (CB) whose band edges relative to the Fermi level (E$_F$) are also indicated by arrows. (b) Right - line map of the scanning tunneling spectrum across the dashed line indicated in the topographic image in the left (V$_b$= -1.00V; I$_T$ = 65pA). (c) Band structure of 3-layer ReS$_2$ and the corresponding DOS.**

We now turn our attention to the band structure of ReS$_2$. Electronically, ReS$_2$ is a semiconductor, however, the nature of the semiconducting gap (direct or indirect) in the bulk versus monolayers remains a subject of discussion both experimentally and theoretically[13,32-39]. A number of experimental techniques including optical spectroscopies and electronic transport found the bulk band gap to be in the range of 1.40 ± 0.07eV [13,15,17,18,40] in agreement with results of theoretical calculations which reported the band gap to be in the range of 1.4 ± 0.1 eV [14,16,19,20,13,34-39]. However, transport and optical spectroscopy are macroscopic measurements which average over large areas of a crystal. Here, we use a local probe, scanning tunneling microscopy and spectroscopy to measure the density of states on the surface of ReS$_2$ crystals. We find that the differential conductance, in Figure 2(a), shows the presence of an energy gap confirming the semiconducting nature of ReS$_2$. The extracted value of the energy gap is 1.35 ± 0.1 $eV$, agreeing with previous experiments. We note that this spectrum, Figure 2(a), represents averages over different samples, areas, and tips as detailed in Supplemental Figure 1 and it is plotted on a logarithmic scale[41], for clarity. In Figure 2(b) we show the spatial variation of the spectrum across device-size area, topographically presented in the left panel by plotting the scanning tunneling spectra across the dashed line (right panel). Within our resolution, the variation of the energy gap is less than 0.1 $eV$. These experimental findings are now compared with our theoretical calculations. In Figure 2(c) we show the band structure and corresponding DOS of a slab of ReS$_2$ consisting of three layers with details of calculations in the supplemental material. The calculations show a gap at the Γ point, and from the corresponding DOS we can estimate the energy gap around E$_{gap}$=1.3 eV, in good agreement with the measured semiconducting energy gap. While the DFT gap appears to be in good agreement with the measured semiconducting energy gap, we note that DFT underestimates the energy gap. Inclusion of many-body corrections at the G0W0 quasiparticle level increases the bandgap to 2.3eV. However, we note that the blue shift of quasiparticle transition energies is partially compensated by excitonic effects which reduces the optical gaps to values often consistent with DFT bandgaps. In a STS measurement, the zero bias corresponds to the Fermi level (E$_F$) as indicated by the dashed line in Figure 2(a). In an electrically neutral ReS$_2$, we expect the position of E$_F$ to be in the middle of the semiconducting gap. However, in our samples we find the position of E$_F$ to be close to the bottom of the conduction band (E$_C$), indicating the crystal is n-doped.

We now turn to the analysis of dopants and defects. When imaging the surface of the ReS$_2$ crystal, we encounter "bright" or "dark" regions representing the presence of a dopant or defect that will electrostatically interact with its environment. Such features have been previously reported on surfaces of doped III-V semiconductors[42], topological insulators[43], transition metal dichalcogenides[44], BP[45,46] or BN[47]. The most common type of defects are shown in the STM topographic image of Figure 3(a). They have a characteristic bright center with a dark halo when imaged at negative bias voltages (Figure 3(b)). As we vary the scanning parameters, we observe that the apparent height of the defects changes, so that they appear bright at negative bias voltages and dark at positive bias voltages. This is demonstrated in Figure 3(c) where we present STM topographic images taken at different bias voltages: -0.80 V, and +0.80 V respectively. The spectroscopic data acquired on the center of the defect, presented in Figure 3(d), closely resembles the one on a defect-free area, and reveals the absence of in-gap states. The slight increase in DOS within the gap could be due to buried defects under the surface[48]. To complement our STM/STS data, we performed XPS analysis of the crystal and, in addition to the expected Re and S we also find the presence of oxygen (Supplemental Figure 4).

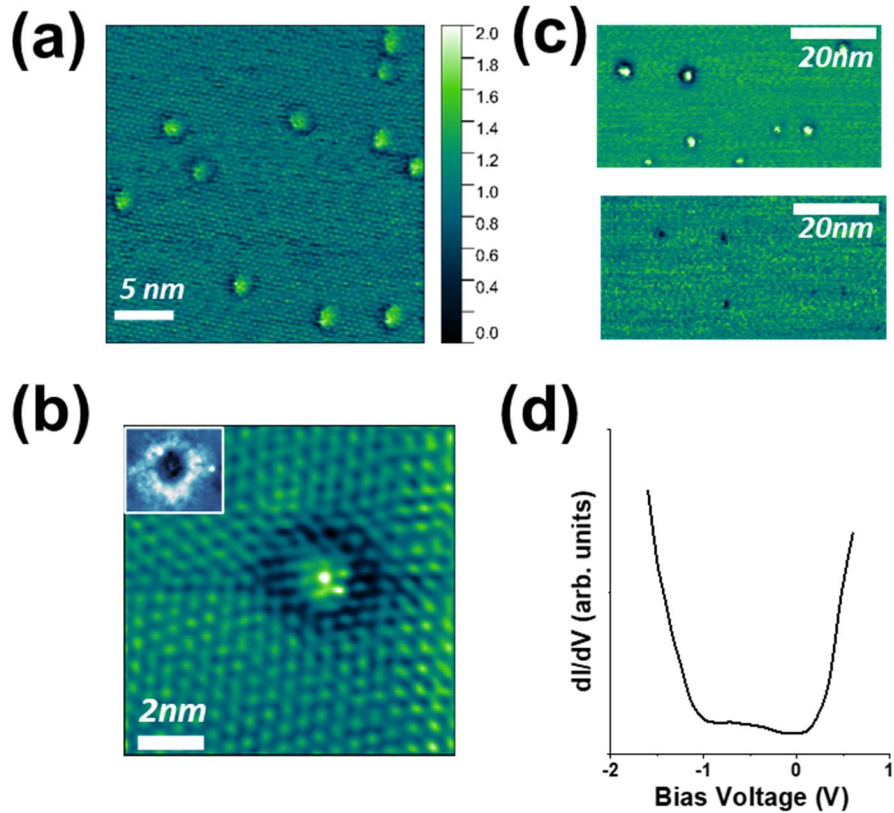

**Figure 3: (a) STM topographic image ($V_b$= -1.20V; $I_T$ = 80pA) showing lattice defects. (b) Topographic high resolution image across a defect. Inset: LDOS map across a defect at $V_B$ = -1.2V and $I_T$ = 35pA. (c) STM topographic images ($I_T$ = 50pA) of an area at different bias voltages. Top: $V_{bias}$= -0.80V; bottom: $V_{bias}$= +0.80 V. (d) Measured STS on the defect in (b).**

Different possible atomic scale defects in the lattice of $ReS_2$ can have signatures in an STM experiment. For example, defects such as a single atom vacancy, single atom adsorption, antisite formation where S atom is substituted by Re atom or vice-versa can be formed during the growth of 2D $ReS_2$ sheets. To elucidate the possible origins of the most common defects observed in our STM experiments on the surface of $ReS_2$ we used DFT calculations. Simulated STM images of pristine $ReS_2$ and defected structures, together with their characteristic density of states, are presented in Figure 4.

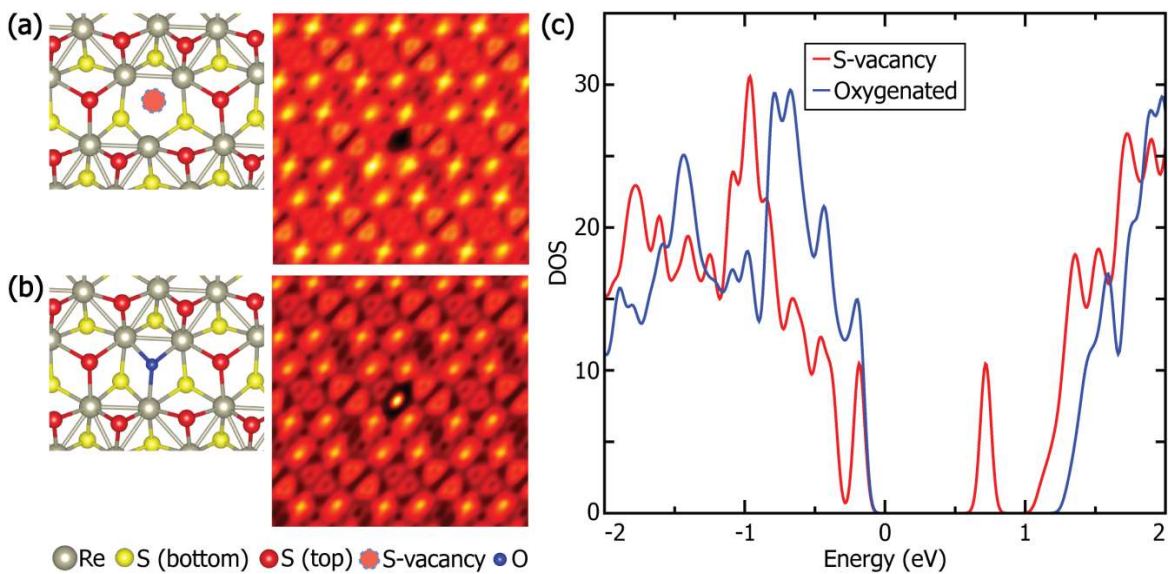

**Figure 4: Crystal structure and simulated STM images of (a) S-vacancy and (b) O absorbed by S-vacancy. (c) LDOS of S-vacancy and O absorbed by S-vacancy.**

The results of ab-initio calculation indictate that Sulfur vacancy, as shown in Figure 4(a), would result in a dark feature in the LDOS, different from the observed common defect. The lack of features characteristic of S vacancy would seem surprising given that previous calculations suggested that sulfur S vacancies have the lowest formation energy[49]. However, we propose that the formation of such defects results in dangling bonds within the lattice, which are readily oxygenated under atmospheric conditions, leading to O atom absorption by vacancies within the lattice. In fact, our theoretical results show that Oxygen absorption by an S-vacancy, as presented in Figure 4(b), will create signatures in the density of states maps that closely resemble the measured defects, with a distinctive halo structure. Moreover, when we examine the density of states, a S-vacancy results in mid-gap states (red curve in Figure 4(c)), while when an O atom is bound to the S vacancy, the semiconducting band gap changes only slightly, in agreement with our experimental finding from STS presented in Figure 3(e), strengthening the case for oxygen being a common defect in the ReS$_2$ layers. These results are also consistent with the reports that oxygen is a source of atomic-scale defects in MoS$_2$[11,12].

We note that in addition to the defects discussed in Figure 3, we also observed less frequent types of lattice defects as discussed in the supplemental material.

In summary, we presented results of scanning tunneling microscopy and spectroscopy together with ab initio DFT calculations of the nanoscale lattice structure, electronic bandgap and defects of in-plane anisotropic semiconducting ReS$_2$. We resolve the chains of rhenium atoms forming diamond-shaped clusters, we measure the semiconducting energy gap and compare the experimental values with results of ab initio calculations. Moreover, we reveal the presence of atomic lattice defects and explore their local properties. By comparing their signatures in the STM/STS experiment with the theoretical calculations, we identify that oxygen is frequently absorbed at defect sites in this material. As the nature of atomic defects is critical in understanding the properties of 2D materials, our result paves the way toward understanding and engineering properties of in-plane anisotropic 2D semiconductors.

## Methods:

**Theory** - The calculations of energy bands, the bandgap and density of states in a single layer and a slab consisting of 3 layers of ReS$_2$ were done using the Quantum Espresso code, with cut-off energy 70 Ry and mesh of 25*25*1 k-grid. Each layer/slab was separated by 20 Å of vacuum and the structure was fully relaxed. Vienna *ab-initio* Simulation Package (VASP) and plane-wave projector-augmented wave (PAW) potentials were used in the calculations.[47] Generalized gradient approximation (GGA) form of Perdew-Burke-Ernzerhof (PBE) was used for the exchange-correlation functional.[48] Kinetic energy cut-off of the plane-wave basis set was taken to be 400 eV. For the structural optimization and LDOS calculations a Gaussian broadening width of 0.05 eV was taken. The convergence criterion between consequent electronic steps was set at $10^{-5}$ eV. Gamma centered k-mesh was taken as 5x5x1 and 10x10x1 for structural optimization and LDOS calculations, respectively, for 2x2x1 supercell. At least 15 Å of vacuum spacing between layers was taken in order to prevent any interactions between layers. STM images were simulated by calculating partial charge densities in the range [-2, 0] and [-2, 2] eV of the cells, and using the formula:

$$E_{total} = \sum_n^h E_n e^{-kz_n}$$

where $E_{total}$ is the summed charge density matrix, n is the layer number, h is the height of the ReS$_2$ monolayer, $E_n$ is the n$^{th}$ layer partial charge density matrix, k is a constant, and $z_n$ is the distance from the artificial STM tip in the z direction. Contrast changes as a results of the bias voltage were simulated by involving electron doped conduction bands in a range of 2 eV.

**STM/STS** We use a commercial RHK Pan Freedom system with ultrahigh vacuum (UHV) and variable temperature capabilities. The data here are taken at temperatures 80K-300K.

**Sample preparation** In this study we use commercial (HQ Graphene) bulk crystals ReS$_2$ cleaved in air and subsequently introduced into ultrahigh vacuum (UHV).

**XPS** The XPS spectra were measured on a Kratos Axis Nova spectrometer equipped with an Al X-ray source. The XPS data were collected using AlKα radiation at 1486.69 eV (150 W, 10 mA), charge neutralizer and a delay-line detector (DLD) consisting of three multi-channel plates.

## Acknowledgement

The authors acknowledge funding from the National Sciences and Engineering Research Council (NSERC) Discovery Grant RGPIN-2016-06717. We also acknowledge the support of the Natural Sciences and Engineering Research Council of Canada (NSERC) through QC2DM Strategic Project STPGP 521420. PH thanks uOttawa Research Chair in Quantum Theory of Materials for support. PP acknowledges partial financial support from National Science Center (NCN), Poland, grant Maestro No. 2014/14/A/ST3/00654 and calculations were performed in the Wrocław Center for Networking and Supercomputing.

H.S. Acknowledges financial support from TUBITAK under the project number 117F095 and from Turkish Academy of Sciences under the GEBIP program. Our computational resources were provided by TUBITAK ULAKBIM, High Performance and Grid Computing Center (TR-Grid e-Infrastructure).

**Supplementary Information for**

**Prevalence of oxygen defects in an in-plane anisotropic transition metal dichalcogenide**

**Plumadore et al.**

1. **Scanning Tunneling spectroscopy**

    Details about the measured variation in the measured band gap across different samples and areas using several tips.

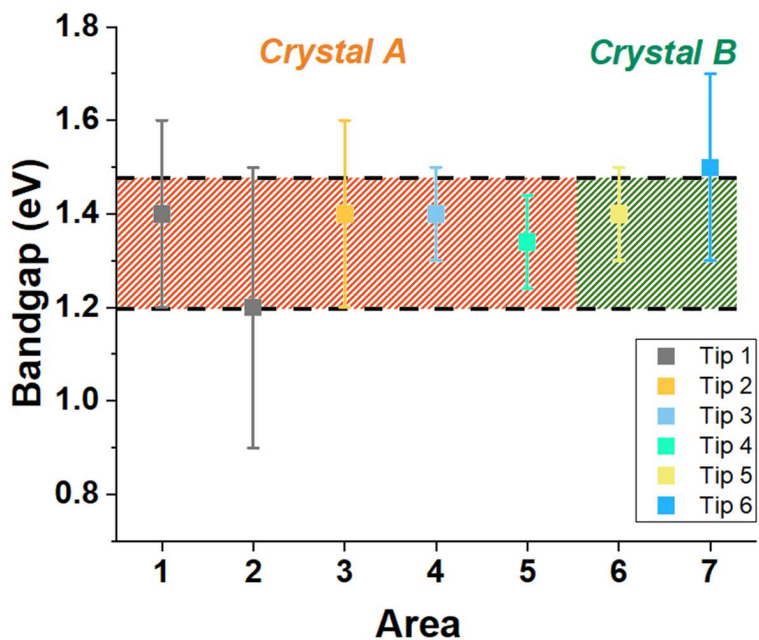

**Supplementary Figure 1: Measured bandgap of the ReS$_2$ for different crystals, areas on a crystal and tips. Error bars reflect uncertainty in identifying the band edges in the STS spectra.**

## 2. Other defects observed in the STM topographic images and LDOS maps

In addition to the common defect described in detail in Figure 3 of the main text (Type A) we also encountered other types of defects indicated in Supplementary Figure 2(a) as Type B and Type C. At negative bias voltages Type B appears bright and Type C appears dark. The relative amount of such defects over an area of 4.7μm$^2$ is presented in Supplementary Figure 2(b) as a histogram, demonstrating that Type A defects are the most commonly present. Supplementary Figure 2(c) shows the change in contrast for Type A defects when the bias voltage changes sign. The characteristics of the dI/dV maps are also different as shown in Supplementary Figure 2(d).

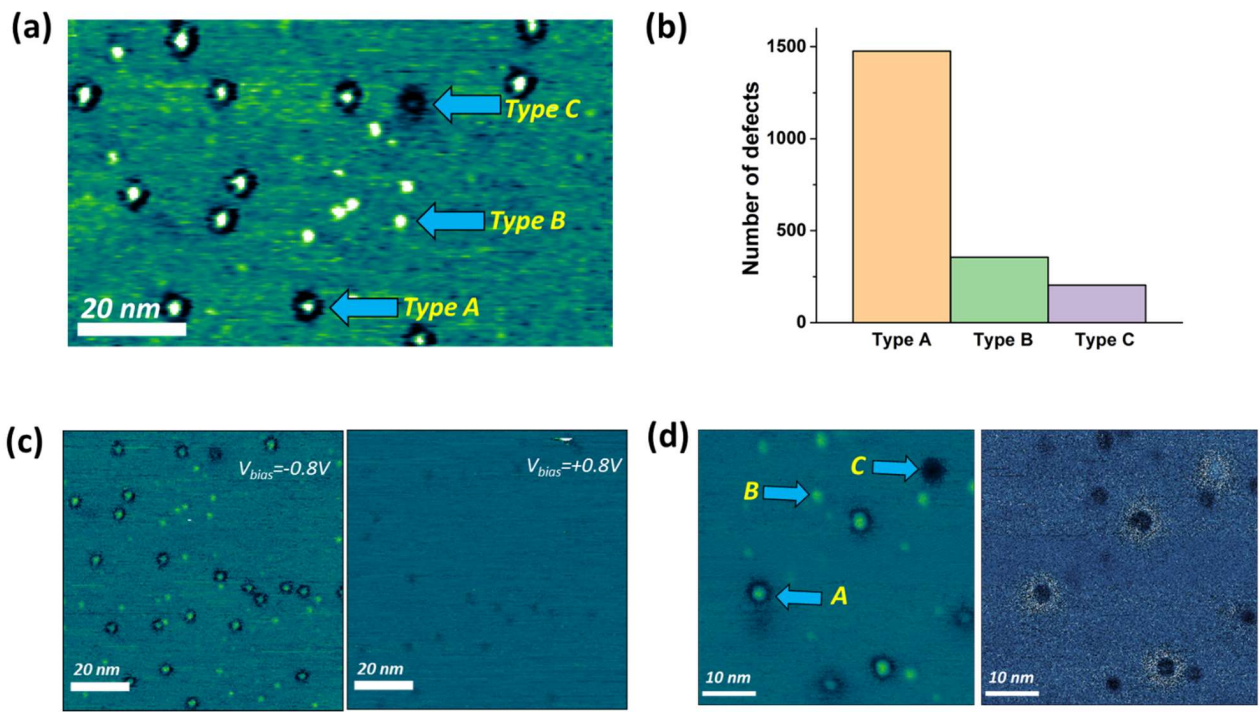

**Supplementary Figure 2: (a) STM topographic image ($V_b$= -1.20V; $I_T$ = 50pA) with different types of defects labeled A, B, C. (b) Histogram of number of defects of each type encountered by scanning a total area of 4.7μm$^2$. (c) STM topographic images ($I_T$ = 50pA) of another area taken at indicated bias voltages. (d) STM topographic image (left) ($V_b$= -0.86 V; $I_T$ =35.0 pA) and its corresponding dI/dV map at -0.86V (right).**

## 3. Details of the theoretical calculation of the signatures of defects

Guided by the signature in the STM images, we explored theoretically the origins of the defects type B and C. In Supplementary Figure 3 we present the results obtained for simulated STM images, calculated LDOS for the indicated type of defects, also illustrated by their lattice structure in the insets. The type of defects that can show bright contrast in the simulated STM images are oxygen adsorbed on pristine ReS$_2$ and Re- antisite presented in Supplementary Figure 3(a) and Supplementary Figure 2(b) respectively. For Type C defects that appear darker in topographic images, the possible origins were identified as S-vacancy, Re-vacancy and S-antisite as shown by Supplementary Figure 3(c)-(e). We note that a similar topographic feature of a bright center with dark halo as type A defect was calculated to correspond to oxygen absorbed on a S-antisite. However, based on the comparison with the spectroscopic feature of Figure 3(e), we can conclude that this is not the origin of type A defect.

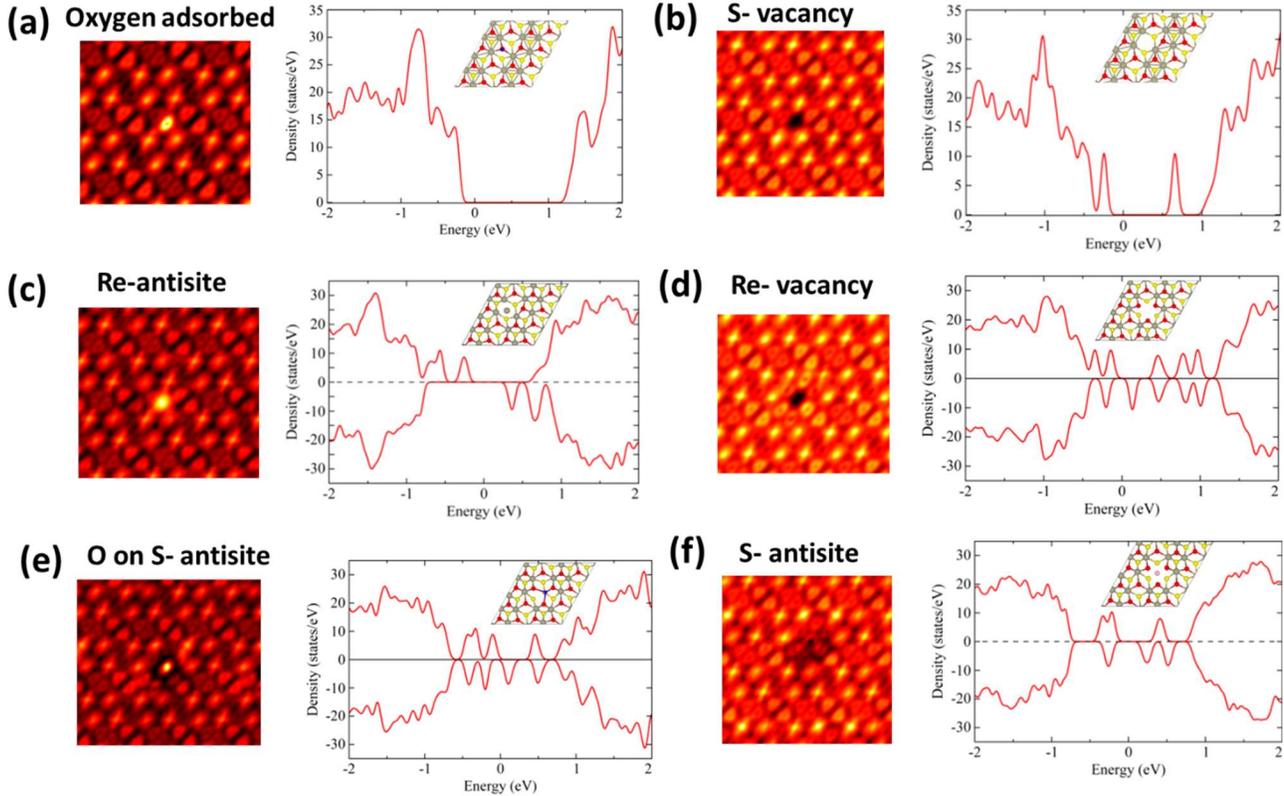

**Supplementary Figure 3: Simulated STM images and LDOS with crystal structure as inset for (a) O adsorbed on pristine ReS$_2$, (b) S-vacancy, (c) Re-antisite, (d) Re-vacancy, (e) O adsorbed on S-antisite (f) S-antisite.**

## 4. Effect of number of layers and stacking on the calculated energy gap

To verify the effect of the number of layers on the energy gap we carry out ab-initio calculations of a slab consisting of 3 layers. In order to obtain the lowest energy stacking configuration, we consider different starting configurations AAA, ABA, ABB and ABC. To ensure that we obtain the lowest energy configuration we follow two relaxation protocols. In the first method, we relax the monolayer $ReS_2$ and let a lattice cell orient in a desired direction in space. One of the consequences is that the "Z lattice parameter" has a small deviation from the "Z axis". We then put another layer above the first layer in AA or AB configuration and relax the cell again. In the third step, we put the third layer in ABA, ABB or AAA sequence and relax the structures. At the end, we find the structure with the lowest energy. We observe that initial ABA stacking changes into ABC stacking after relaxation which turns out to have the lowest energy.

In the second method, we first relax the monolayer and freeze "Z lattice parameter" direction along Z axis. We make AA and AB stacking, and create bulk crystal and relaxed it. The lattice constant of bulk in Z direction increases to avoid interaction with the next nearest unit-cells. From such relaxed bulk, we take AB stacking bilayer and add another layer above, making ABA stacking of 3 layers. After relaxation we obtain the total energy which is nearly 15 meV/chemical-unit lower as compared with the energy obtained from the first method. However, the energy difference between the two different stackings of 3 layers (ABA and ABC) is small and we do not exclude a mixture of them at room temperature. In Fig. 2(c) we show the band structure for ABA stacked 3 layer $ReS_2$ for wave-vector k from M to Γ to K. We see that the valence band maximum is at the Γ point with direct Kohn-Sham energy bandgap between CB and VB of 1.3 eV. However, stacking of layers as we move away from the surface inside the bulk and how the stacking affects the bandgap is not clear. The bulk and single layer bandgaps were found to be very similar which was one of the arguments for the layer independent properties of $ReS_2$. We note however that the peak in the density of states at the edge of the valence band disappears for the 3 layer system which may explain the absence of the experimental observation of this sharp peak. However, the absence of the peak may be related to the presence of impurities. Additional studies are needed to improve agreement between experiment and theory but the measured and calculated bandgap of $ReS_2$ ~1.4eV agree very well.

## 5. Further details of the XPS characterization

The XPS spectra were measured on a Kratos Axis Nova spectrometer equipped with an Al X-ray source. The XPS data were collected using AlKα radiation at 1486.69 eV (150 W, 10 mA), charge neutralizer and a delay-line detector (DLD) consisting of three multi-channel plates.

Upon analysis, we find the Re 4f core level spectra positions at 42.6eV (4f 7/2) and 45 eV (4f 5/2), consistent with literature for exfoliated $ReS_2$ [1,2]. The S 2p core level region can be fit by two S 2p 3/2 and 1/2 doublets (163.3eV, 164.5eV) and (162.6, 163.8eV) as reported previously and attributed to the 1T-$ReS_2$ structure[1,2]. In addition, we also find the presence of oxygen, consistent with our interpretation of the measured atomic defects (Supplementary Figure 4).

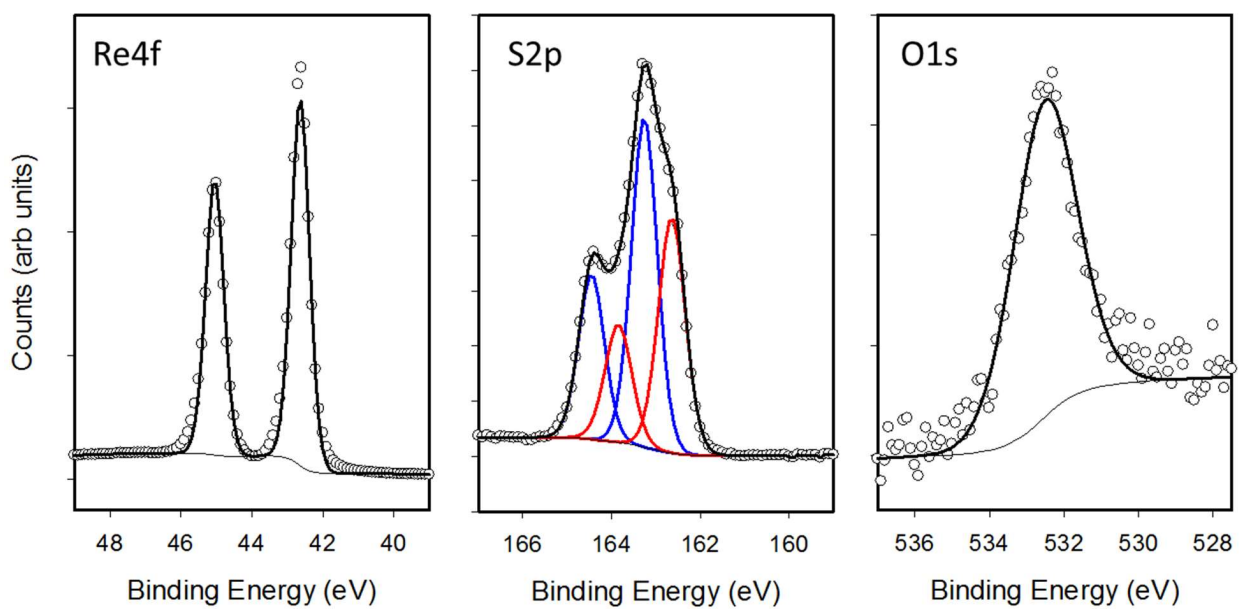

**Supplementary Figure 4: High resolution XPS spectra of Re4f, S2p and O1s regions**

**Supplementary references**

1. Khosravi et al., Materials 2019, 12, 1056
2. Fujita et al., Nanoscale 2014, 6, 12458